# Establishing quasi-linear quadrupole functional topology by oxygen-vacancy engineering at a ferroelectric domain wall

Hemaprabha Elangovan,[†,1,2] Maya Barzilay,[†,1,2] Jiawei Huang,[3,4,5] Shi Liu,[3,4,5] Shai Cohen[6] and Yachin Ivry[1,2,*]

[1] Department of Materials Science and Engineering, Technion – Israel Institute of Technology, Haifa 3200003, Israel.

[2] Solid State Institute, Technion – Israel Institute of Technology, Haifa 3200003, Israel.

[3] School of Science, Westlake University, Hangzhou, Zhejiang 310024, China.

[4] Institute of Natural Sciences, Westlake Institute for Advanced Study, Hangzhou, Zhejiang 310024, China.

[5] Key Laboratory for Quantum Materials of Zhejiang Province, Hangzhou Zhejiang 310024, China.

[6] Nuclear Research Centre-Negev, Beer-Sheva 84190, Israel.

[*]Correspondence to: ivry@technion.ac.il

[†] These authors contributed equally to the work.

**Abstract**

Oxygen vacancies in two-dimensional metal-oxide structures garner much attention due to unique conductive, magnetic and even superconductive functionalities they induce. Ferroelectric domain walls have been a prominent recent example because they serve as a hub for topological defects that enable unusual symmetries and are relevant for low-energy switching technologies. However, owing to the light weight of oxygen atoms and localized effects of their vacancies, the atomic-scale electrical and mechanical influence of oxygen vacancies has remained elusive. Here, stable individual oxygen vacancies were found and engineered *in situ* at domain walls of seminal titanate perovskite ferroics. The atomic-scale strain, electric-field, charge and dipole-moment distribution around these vacancies were characterized by combining advanced transmission electron microscopy and first-principle methodologies. 3-5 % tensile strain was observed at the immediate surrounding unit cells of the vacancies. The dipole-moment distribution around the vacancy was found to be an alternating head-to-head – tail-to-tail – head-to-head structure, giving rise to a quasi-linear quadrupole topology. Reduction of the nearby Ti ion as well as enhanced charging and electric-field concentration near the vacancy confirmed the quadrupole structure and illustrated its local effects on the electrical and structural properties. Significant intra-band states were found in the unit cell of the vacancies, proposing a meaningful domain-wall conductivity. Oxygen-vacancy engineering and controllable quadrupoles that enable pre-determining both atomic-scale and global functional properties offer a promising platform of electro-mechanical topological solitons and device miniaturization in metal oxides.

**Introduction**

Oxygen vacancies in metal-oxides are known to have strong effects on the material functional properties, hence garnering much attention.[1] Examples include ferromagnetism in LaCoO$_3$,[2] conductive filament formation in metal-oxide-metal structures,[3] enhanced catalytic activity in SrCoO$_{3-\delta}$,[4,5] and ferroelectric-superconductivity coexistence in Sr$_{1-x}$Ca$_x$TiO$_{3-\delta}$,[6] even though these materials are insulating and non-magnetic in their stoichiometric form. Recent focus is put on oxygen vacancies in 2D structures. Interface between two metal oxides, such as LaAlO$_3$/SrTiO$_3$ is a common example.[7–9] A prominent platform for such 2D structures is ferroelectric domain walls, which separate neighboring regions with different dipole-moment and polarization orientations.[10] Oxygen vacancies have lately been believed as serving as a hub for topological defects, such as vortices and skyrmions as well as altering significantly the properties of domain walls,[11–13,14] including introduction of magnetism in Hf$_{0.5}$Zr$_{0.5}$O$_2$[15] and superconductivity in WO$_x$[16] that are restricted to the 2D domain-walls, which in turn are encapsulated within the high-dielectric parent ferroic material. Substantial effort has been given though to domain-wall conductivity in large-bandgap ferroelectrics, vastly because domain walls are movable by will with external electric fields, giving rise to miniaturized memristive cells.[17–19] Attention has been given mainly to BiFeO$_3$,[20–23] ErMnO$_3$[24,25] and LiNbO$_3$[26–28] that show high and reproducible conductivity, but the success of presenting domain-wall conductivity in the traditional perovskite ferroelectrics BaTiO$_3$[29] and Pb(Zr$_{0.2}$Ti$_{0.8}$)O$_3$ has remained limited.[30,31]

Macroscopic-scale effects of oxygen vacancies are observed readily as an averaged behavior with spectroscopic methods.[32–36] Changes in the material electronic band structure and charge-carrier concentration, both contributing to conductivity, have been observed as coupled with an increase in oxygen-vacancy concentration.[37,38] At the mesoscopic scale, enhanced conductivity is frequently observed at walls that separate head-to-head (*h-h*) or tail-to-tail (*t-t*) polarization domains with scanning probe microscopy.[39] Theoretical[40] and modeling[41] analyses of these structures predict that these charged *h-h* (*t-t*) domain walls repel (or attract) oxygen vacancies, which in turn help stabilize these charged domain walls globally. Yet, basic questions, including what the exact distribution of strain, electric charge, electric field and dipole moments around a vacancy is, have remained open. Recent atomic-scale scanning transmission electron-microscopy (STEM) studies of bismuth ferrite demonstrated clusters of charged point defects at domain walls.[42,43] Initially, indirect observations attributed these defects to oxygen vacancies.[44,45] However, very lately, electron energy loss spectroscopy (EELS) studies determined that charged defects at the domain walls are most likely due to bismuth vacancies, questioning the existence and role of oxygen vacancies at the domain walls.[46] A major reason for the unresolved role of oxygen vacancies is that modeling point defects at domain walls within a collectively-interacting-dipole matrix is complex. Likewise, observing these light atoms is a non-trivial task, and spotting individual vacant sites is even more challenging.[47,48] The mechanisms causing these effects have also remained elusive due to difficulties in characterizing and modeling individual oxygen vacancies in general and more so even at the domain wall.[49]

Here, we combined various atomic-scale imaging and manipulation techniques together with DFT modeling to demonstrate the formation of oxygen vacancies at domain walls in the seminal ferroelectric perovskite $BaTiO_3$. Our results reveal that an individual vacancy induces mechanical strain and electrical charging at a distance of up to two unit cells, which is much smaller than previous predictions. Dynamic measurements showed that upon the disappearance of the vacancy, the domain wall becomes unstrained and uncharged. Complementary DFT calculations expanded the work also to other perovskite ferroelectrics and not only supported the stability of oxygen vacancies at the domain walls, but also illustrated that they give rise to intragap states, which are significant for conductance. Lastly, careful analysis of the dipole-moment distribution around the vacancy showed a quasi-linear quadrupole that comprises a pair of *h-h* and *t-t* dipole moments. The contribution of this nonconventional topological structure to the local electric field distribution at the domain wall was demonstrated independently.

**Results and discussions:**

Single-crystal $BaTiO_3$ 50-nm crystallites were used, providing strain-free bulk-like behavior, while yet allowing sufficient electron transparency to aid the TEM analyses. Recently, it has been shown that ferroic domain walls can be formed, moved and switched contactless in such materials *in-situ* during atomic-scale TEM imaging.[50,51] We used this method to form the domain wall structure, which served as a template for the oxygen vacancies. Simultaneous STEM-based differential phase contrast (DPC), integrated differential phase contrast (iDPC) high-angle annular dark field HAADF and EELS where used to characterize the atomic-scale chemical-element, strain, charge and electric-field distribution, exact atomic location of the ions and ion's oxidation state, respectively. A key factor that helped with the characterization is that while the contrast in traditional STEM and HAADF methods is quadratic with the atomic number--decreasing the signal-to-noise ratio of light elements--the DPC methods are sensitive also to lighter elements, such as oxygen.[52] The contrast in DPC and iDPC can be also more localized at a specific focal plane than conventional STEM imaging methods. Hence, by changing the focal depth, tomography-like characterization of the structural and electric properties was obtained. Moreover, in DPC, the electron beam diversion due to local changes in the electrical and magnetic fields within the sample are detected, allowing us to extract the local electric-field distribution (see detailed discussion in the SI). Lastly, the high accuracy of atom location obtained with iDPC[52] was used to map the dipole-moment distribution with confidence.

Figure 1A shows the atomic structure (iDPC) of an area in which artificial domains (stiped 90° domains) where formed *in situ*. Oxygen vacancies appear clearly as points of 'missing' atoms with black contrast. The vacancies are located along the domain walls that appear not only in Figure 1A, but also in the simultaneously imaged electric-field distribution in that area (DPC, Figure 1B) and HAADF image (Figure S1).

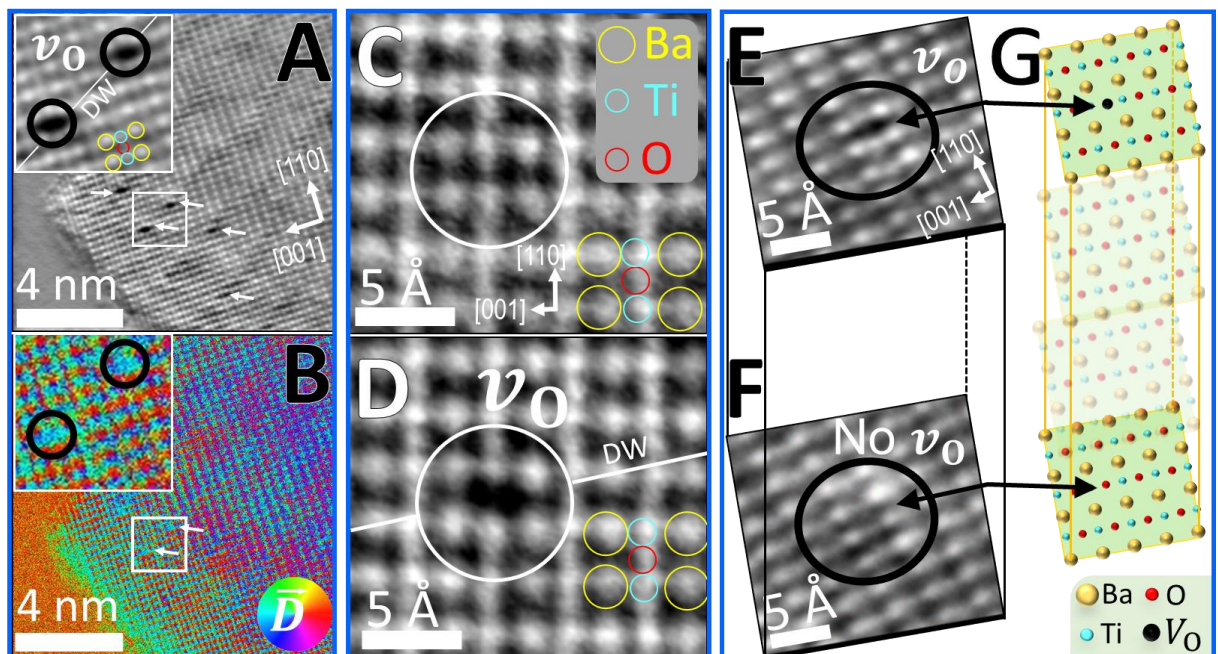

**Figure 1| Formation and direct observation of oxygen vacancies at 90° domain walls in BaTiO$_3$.** (**A**) Atomic structure (iDPC image) and (**B**) the simultaneously imaged electric-field distribution at the same area (DPC micrograph) of artificially formed oxygen vacancies at 90° domain walls in BaTiO$_3$ crystallite. Representative oxygen vacancies ($v_O$) are highlighted in the insets. Striped domain walls (DW) appear as change in contrast in (A), (B) and the simultaneously imaged HAADF signal (Figure S1). Color wheel in (B) represents the orientation (color) and intensity (hue) of the electric-field displacement vector. (**C**) Atomic structure of an unperturbed BaTiO$_3$ crystallite. (**D**) The same area after intentional oxygen-vacancy formation (exposure to 0.17 pA/pm$^2$ *in situ*). (**E**) An oxygen vacancy at a certain depth within a BaTiO$_3$ crystallite. (**F**) Changing the focal depth to a plane that is located 2±1 nm exactly below (E) shows that all the atomic sites are occupied and no oxygen vacancies exist. (**G**) Schematic depth profiling of the oxygen-vacancy localization as was observed in (E-F). Larger scale images of (C-F) are given in Figure S1.

To show controllability of the vacancies, first, the atomic structure of an unperturbed crystal was mapped (Figure 1C). Figure 1D shows the same area after successful intentional formation *in situ* of an oxygen vacancy within a domain wall (details about the domain-wall formation procedureare provided in Reference[50]).

Localization of the oxygen vacancies was demonstrated with the aid of high sensitivity to focal depth of the iDPC method.[53] Figures 1E-F show that the vacant sites are localized at a certain plane within the material. That is, although the vacancy is noticeable in Figure 1E, by changing carefully the focal depth, Figure 1F shows that the same site at the same area albeit at a plane that is placed only 2±1 nm below the plane in Figure 1F is now occupied by an oxygen atom and no vacancy was detected (see schematics in Figure 1G). Large-scale micrographs of Figures 1A-F are given in Figure S1.

Figure 2A shows the dipole-moment distribution across the domain wall based on the Ti ion displacement (the domain wall is identified also with a simple contrast analysis of the larger scale HAADF, DPC and iDPC signals, see Figure S2). Two oxygen vacancies were observed here along the wall and the dipole-moment distribution around them was identical. In both cases, the dipole moments were aligned *t-t* at the site with the missing oxygen atom, accompanied by a *h-h* alignment at the immediate neighboring unit cells from each side. Beyond the distance of already one unit cell, the dipole moments returned to their unperturbed *h-t* orientation. Previous studies proposed that charged vertices of dipole moments (non-*h-t*) are accompanied with strain release.[38] Figure 2B shows the strain distribution across the unit cells with charged vertices (top), which was evaluated based on the distance between neighboring barium atoms. First, 287±5 pm Ba-Ba distance was measured for unit cells with no oxygen vacancies around them, which is in agreement with the literature.[54] In comparison to this value, Figure 2B (bottom) shows 5% expansion for the vacant site (302±3 pm) and 3.5% expansion for the immediate neighboring sites (297±2 pm). The strain was relaxed already at the next unit cells, complying with the dipole moment distribution. To show the reproducibility of this behavior, data was collected for three oxygen-vacancy sites and over fifty unperturbed unit cells as demonstrated in Figure S3.

To further confirm the charging at the vacant site, atomic-resolution EELS measurements were done across the domain wall. Figures 2C and S3 reveal that the oxidation state of the titanium ion changes from $Ti^{4+}$ within the domains to a lower oxidation state at the domain wall,[55–58] where the oxygen vacancies are.

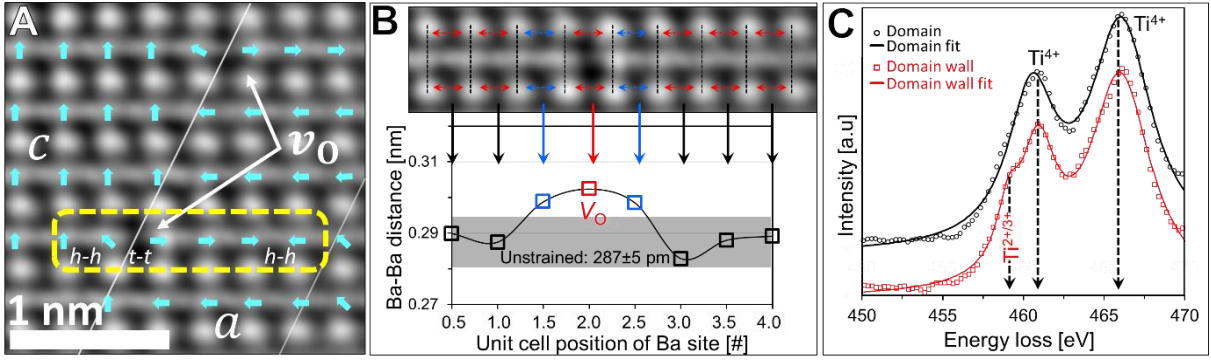

**Figure 2 | Atomic-scale mapping of the charge-strain distribution near an oxygen vacancy site at a domain wall.** (**A**) An iDPC image of oxygen vacancies at a domain wall. A quasi-linear quadrupole (*h-h t-t h-h*) dipole-moment arrangement around a vacancy is marked in green dash lines. (**B**) The Ba-Ba distance distribution near the vacancy demonstrates 5% tensile strain at the vacant site, and 3.5% at the adjacent site. The following unit are unstrained. (**C**) EELS spectra from domains and domain wall near the Ti-$L_{23}$ peak, showing a reduced state of titanium ions at the domain walls (Figure S4).[55–58]

Summing up the dipole-moment distribution around the oxygen vacancy shows a *h-h – t-t – h-h* structure. This alternating dipole moment organization represents an individual unit of a quadrupole. Because the dipole moments near one of the *h-h* vertices has an additional orthogonal component we can refer to this structure as a quasi-linear quadrupole. Traditionally, dipole moments dominate the behavior of ferroelectrics and other dielectric materials. Higher orders of the multipole expansion, such as quadrupoles, are considered typically as negligible for the macroscopic behavior of the material because their contribution is significant only at the short range.[59] Nevertheless, in the context of domain walls, and in particular oxygen vacancies and other point defects at domain walls, short-range interactions become prominent.[60]

Thus, next, the short-range electric-field distribution around oxygen vacancies was examined at sub-atomic scale by means of DPC. Figure 3A shows the atomic structure of a $BaTiO_3$ crystallite with an oxygen vacancy at the domain wall. The simultaneous iDPC mapping in Figure 3B shows the absence of an electric field at the vacancy site. Likewise, a locally enhanced electric field is observed around the titanium ion that is at the adjacent *h-h* charged vertex (note that to reduce the noise level, 5 × 5 pixel averaging was done[44,61]). This local electric-field enhancement is typical for such a quasi-linear quadrupole.[62] Following Gauss's law, the local charge distribution can be extracted from the electric field; however, this should be done carefully. Mostly, the local field that is mapped with DPC is the electric field, $E$, which corresponds to the free charge density, $\rho_f$. Because in ferroelectrics, bound charge ($\rho_b$) and the resultant polarization dominate $\rho_f$ and $E$, the entire displacement field ($D$) must be taken into account[63] (see additional technical discussion in the SI). That is, the charge density around the oxygen vacancy can be extracted from taking the divergence of the DPC mapping (Figure 3B):

$$\nabla \cdot D \approx \frac{\rho_t}{\varepsilon \varepsilon_0} \quad (1),$$

where $\rho_t$ is the total bound and free charge density, $\varepsilon_0$ is vacuum permittivity and $\varepsilon$ is the dielectric constant of the material. DPC micrographs are two-dimensional. Thus, integration of Equation 1 over the sample thickness allows us to extract the areal charge density ($\sigma_t$).

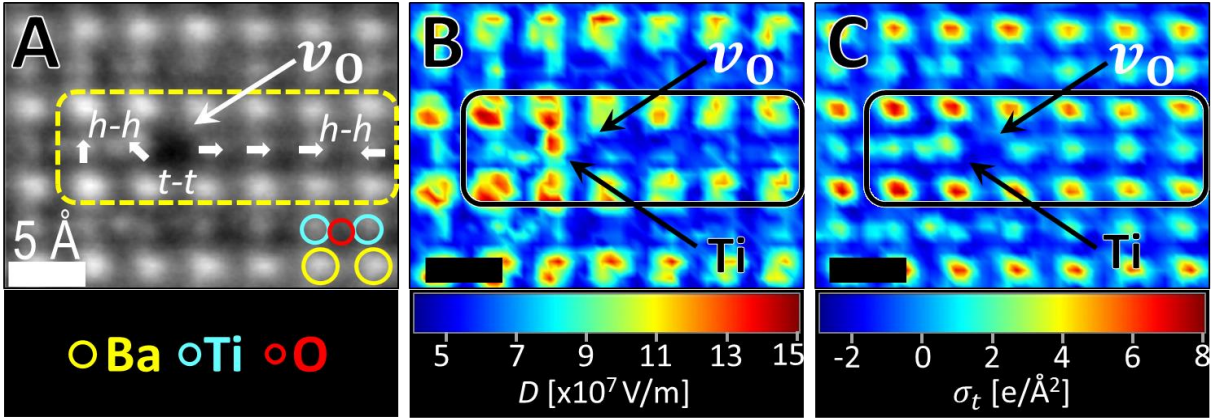

**Figure 3| Enhanced displacement field and charge density near an oxygen vacancy. (A)** iDPC image near an oxygen vacancy (a quasi-linear quadrupole is highlighted). **(B)** Displacement-field ($D$) distribution around the vacancy showing electric-field enhancement at the *h-h* dipole near the vacant site. **(C)** Calculated (Equation 1) charge density around the oxygen vacancy, showing only background charge at the vacant site as well as a strong charge of a larger ionic radius for the nearby Ti ion (both are highlighted). All scale bars are 5Å. Larger scale images are given in Figure S5.

Figure 3C shows the charge distribution around the oxygen vacancy. Dominant charge density is observed at the Ba sites with respect to Ti and oxygen sites everywhere. Only background charge is observed at the vacant site, whereas the charge at the Ti site near the region with the higher displacement field is distributed over a large area with respect to the other titanium ions. Because the atomic radius of a titanium ion increases with decreasing oxidation states,[64] the latter is in agreement with the above EELS result. Note that the oxidation state of the titanium ion was changed only in the charged *h-h* vertex closest to the vacancy. This asymmetric charge distribution is accompanied by an off-axis diagonal displacement of the reduced Ti ion (Figure 3C).

There is a strong motivation to understand the effects of oxygen vacancies on the global material behavior of the ferroelectric, in addition to the above characterized local microscopic effects. In particular, there is a technological interest in understanding the effects of oxygen vacancies on domain-wall conductivity. Thus, complementary modelling was done to expand the realm of the above microscopic characterization. Figure 4A shows a density functional theory (DFT) modeling of a 90° domain wall in BaTiO$_3$ that comprises an oxygen vacancy.

We note that modeling a 90° domain wall in the tetragonal phase of BaTiO$_3$ with DFT turns out to be a nontrivial task. First, DFT calculations are typically done at 0 K at which BaTiO$_3$ adopts the rhombohedral ground-state phase, whereas a tetragonal structure of BaTiO$_3$ is stable only at finite temperatures. Second, the energy of a 90° domain wall in BaTiO$_3$ is expected to be even smaller than that of a 180° domain wall (7.84 mJ/m$^2$).[65,66] In comparison, the energies of 180° and 90° walls in PbTiO$_3$ are 170 mJ/m$^2$ and 90 mJ/m$^2$, respectively.[67] Modeling a low-energy interface demands a high accuracy in energy and atomic forces. Moreover, to maintain a reasonable oxygen vacancy concentration at the wall, a relatively large supercell is needed, further increasing the computational cost. These subtle issues likely explain the lack of DFT studies on 90° domain walls in BaTiO$_3$ in literature. Here we followed a protocol developed to study highly unstable charged domain walls.[41] A $10\sqrt{2} \times 2\sqrt{2} \times 1$ supercell of 200 atoms was divided into three zones: atoms in the two end zones were fixed to the bulk values of the tetragonal phase while atoms in the middle region were allowed to relax. This allowed for structural optimization of a low-energy 90° wall sandwiched by two bulk tetragonal domains (fixed end zones). First, the optimized structure of defect-free 90° walls was obtained (Figure 4A) and then an oxygen vacancy was introduced at the wall, which was located at the center of the supercell (Figure 4B). The local polarization of each unit cell was evaluated by the displacement of the central Ti atom with respect to its surrounding oxygen octahedron.

The DFT simulations agree with the experimental data for several important aspects. First, the simulations show an asymmetric dipolar distribution of the charged *t-t* and *h-h* vertices near the oxygen vacancy. Likewise, oxygen vacancies are found stable also at domain walls that are as dense as 2 nm periodicity. Finally, Figure 4B shows that the dipole moments relax already at the distance of ca. 2 unit cells from the vacancy.

To examine the effect of oxygen vacancies on the domain-wall conductivity, layer-resolved local density of states (LDOS) was computed. Figure 4C shows the existence of significant intra-gap states at the vacant site (fewer states exist also at the nearby unit cells), whereas no mid-gap states were observed elsewhere. As a comparison, we examined the effects of oxygen vacancy on the electronic structures of 90° domain walls in PbTiO$_3$. Three different types of oxygen vacancies at the wall were studied, all leading to meaningful intra-gap states at the interface (Figure S6). This suggests that the defect levels due to oxygen vacancies are promising for enhanced conductivity.

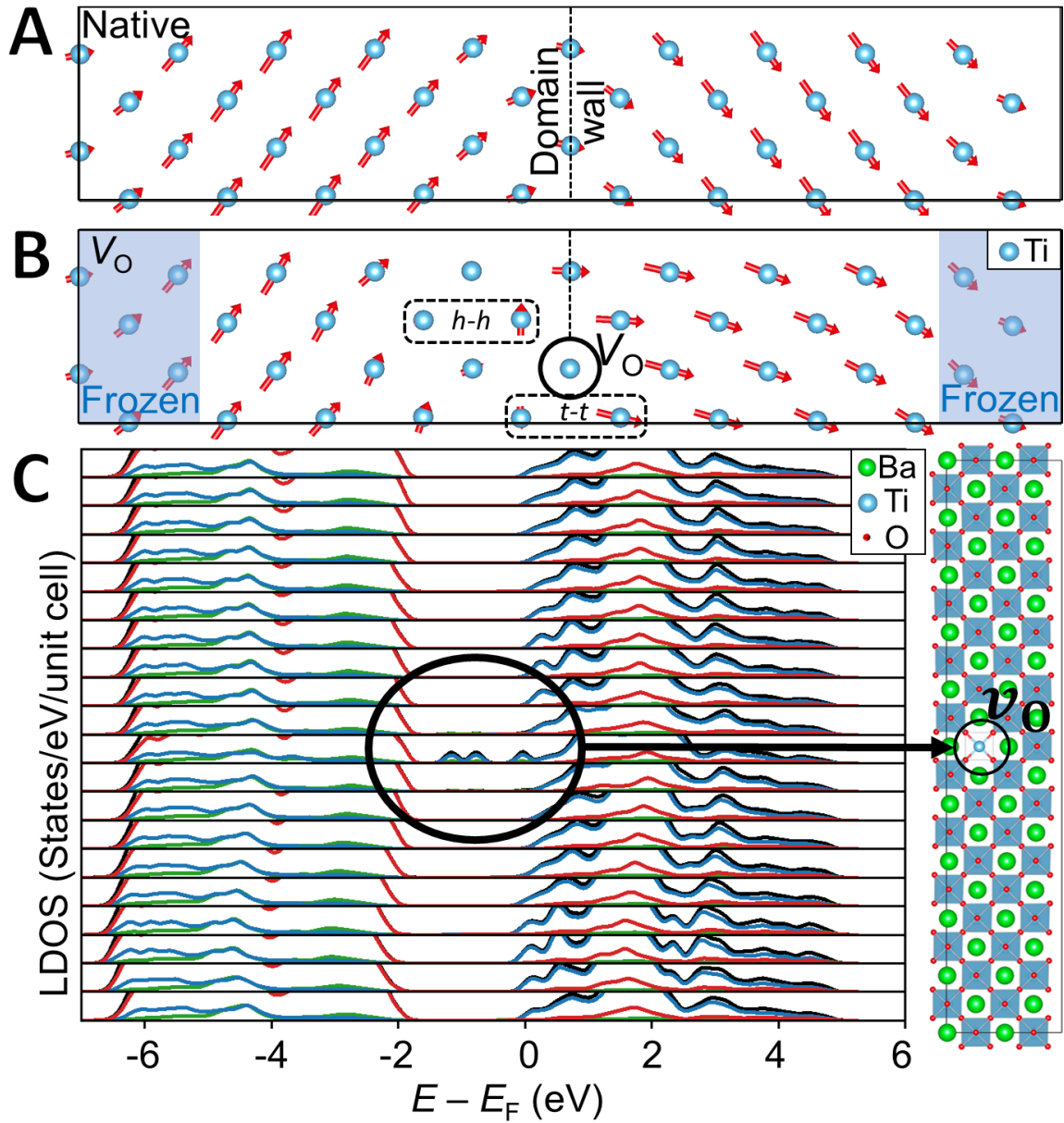

**Figure 4| Local structure and projected density of states around an oxygen vacancy computed with DFT.** (**A**) Dipole-moment distribution near 90° domain wall in BaTiO$_3$. (**B**) A stable oxygen vacancy at the domain wall with an induced charging *h-h* and *t-t* vertices. (**C**) Contribution to the local DOS (Left) of individual unit cells near an oxygen vacancy at a 90° domain wall (Right). Red, blue, and green curves represent contributions from O, Ti, and Ba atoms, respectively, while the black curve is the total DOS.

**Conclusions**

Most studies propose that domain-wall conductivity arises from a coherent *h-h* or *t-t* dipole-moment organization at the domain wall, which in turn can serve as a hub for oxygen vacancies (and other point defects). The above results show that *h-h* and *t-t* can be local effects, giving rise to substantial changes in the local electric-field, charge, mechanical strain, dipole and quadrupole moment distribution at the domain wall as well as in the oxidation state of nearby ions. It is possible that local effects bare more prominent in the case of 90° ferroelastic domain walls than in 180° domain walls because of the lower symmetry of the former. However, we believe that future work is required to examine this hypothesis.

A quasi-uniaxial quadrupole that is located at a domain wall which separates between two areas with clear dipole-moment and polarization orientation. This spatially localized change in symmetry is similar to a sudden change in direction in the middle of a spiraling telephone cord indicating that the quadrupoles in this work serves as a topological soliton. This topological structure is promising for nanoscale electronic technologies and deserves further investigation.

The contribution of an atomic-scale topological structure is found important not only to local properties but also for macroscopic properties, such as electric conductivity. The above results highlight the significance of local topological and structural effects, such as quadrupoles, which have been considered negligible with respect to the macroscopic and device-relevant properties. Likewise, the above oxygen-vacancy engineering methodologies can be implemented for manipulating exotic structures and functional properties of ferroelectrics and other metal-oxide materials.

**Methods:**

Material: Commercially available single crystal $BaTiO_3$ nanoparticles of 50 nm size[50,55,68] were purchased from US Research Nanomaterials, Inc (99.9% pure). To evenly spread the particles on the TEM grid, the particles were suspended in ethanol and sprayed on the amorphous carbon coated holey Cu grid using nitrogen gas.

TEM: The TEM experiments were carried out on an aberration-corrected Titan Themis 80–300 operated at 200 kV. For STEM-DPC experiments, the equipment was set with a dose of 20 – 250 pA. The probe semi-convergence angle of 21 mrad and collecting semi-angles of 25–154 mrad was used for HAADF mode. The DPC images were captured using segmented (4-quadrant) annular dark-field detector (DF4), with collecting semi-angle of 6–34 mrad.

DFT: All first-principle DFT calculations were performed using QUANTUM ESPRESSO[69] with generalized gradient approximation of the Perdew-Burke-ErnZerhof for solids (PBEsol) type. Given the large supercell used to model domain walls, we used GBRV ultrasoft pseudopotentials[70] and a $1 \times 1 \times 4$ Monkhorst-Pack $k$-point grid for structural optimization and a $2 \times 2 \times 8$ $k$-point grid for electronic structural calculation. The plane-wave cutoff is set to 40 Ry and the charge density cutoff is set to 200 Ry, respectively.

**Acknowledgements**

The Technion team acknowledges support from the Zuckerman STEM Leadership Program, the Technion Russel Barry Nanoscience Institute, Pazy Research Foundation grant #149-2020 and the Israel Science Foundation (ISF) Grant No. 1602/17. We also thank Dr. Yaron Kauffman and Mr. Michael Kalina for technical support. J.H. and S.L. acknowledge the supports from Westlake Education Foundation. The computational resource is provided by Westlake HPC Center.

**References**

(1) Gunkel, F.; Christensen, D. V.; Chen, Y. Z.; Pryds, N. Oxygen Vacancies: The (in)Visible Friend of Oxide Electronics. *Appl. Phys. Lett.* **2020**, *116* (12), 120505.

(2) Yan, J. Q.; Zhou, J. S.; Goodenough, J. B. Ferromagnetism in $LaCoO_3$. *Phys. Rev. B* **2004**, *70*, 014402.

(3) Ota, T.; Kizaki, H.; Morikawa, Y. Mechanistic Analysis of Oxygen Vacancy Formation and Ionic Transport in $Sr_3Fe_2O_{7-\delta}$. *J. Phys. Chem. C* **2018**, *122* (8), 4172–4181.

(4) Jeen, H.; Choi, Woo S.; Biegalski, M. D.; Folkman, C. M.; Tung, I. C.; Fong, D. D.; Freeland, J. W.; Shin, D.; Ohta, H.; Chisholm, M. F.; Lee, H. N. Reversible Redox Reactions in an Epitaxially Stabilized $SrCoO_x$ Oxygen Sponge. *Nat. Mater.* **2013**, *12* (11), 1057–1063.

(5) Petrie, Jonathan R.; Jeen, Hyoungjeen; Barron, Sara C.; Meyer, Tricia L.; Lee, Ho Nyung. Enhancing Perovskite Electrocatalysis through Strain Tuning of the Oxygen Deficiency. *J. Am. Chem. Soc.* **2016**, *138* (23), 7252–7255.

(6) Rischau, C. W.; Lin, X.; Grams, C. P.; Finck, D.; Harms, S.; Engelmayer, J.; Lorenz, T.; Gallais, Y.; Fauqué, B.; Hemberger, J.; Behnia, K. A Ferroelectric Quantum Phase Transition inside the Superconducting Dome of $Sr_{1-x}Ca_xTiO_{3-\delta}$. *Nat. Phys.* **2017**, *13* (7), 643–648.

(7) Ohtomo, A.; Hwang, H. Y. A High-Mobility Electron Gas at the $LaAlO_3/SrTiO_3$ Heterointerface. *Nature* **2004**, *427* (6973), 423–426.

(8) Salluzzo, M.; Gariglio, S.; Stornaiuolo, D.; Sessi, V.; Rusponi, S.; Piamonteze, C.; De Luca, G. M.; Minola, M.; Marré, D.; Gadaleta, A.; Brune, H.; Nolting, F.; Brookes, N. B.; Ghiringhelli, G. Origin of Interface Magnetism in $BiMnO_3/SrTiO_3$ and $LaAlO_3/SrTiO_3$ Heterostructures. *Phys. Rev. Lett.* **2013**, *111* (8), 087204.

(9) Bert, J. A.; Kalisky, B.; Bell, C.; Kim, M.; Hikita, Y.; Hwang, H. Y.; Moler, K. A. Direct Imaging of the Coexistence of Ferromagnetism and Superconductivity at the $LaAlO_3/SrTiO_3$ Interface. *Nat. Phys.* **2011**, *7* (10), 767–771.

(10) Nataf, G. F.; Guennou, M.; Gregg, J. M.; Meier, D.; Hlinka, J.; Salje, E. K. H.; Kreisel, J. Domain-Wall Engineering and Topological Defects in Ferroelectric and Ferroelastic Materials. *Nat. Rev. Phys.* **2020**, *2* (11), 634–648.

(11) Seidel, J.; Vasudevan, R. K.; Valanoor, N. Topological Structures in Multiferroics - Domain Walls, Skyrmions and Vortices. *Adv. Electron. Mater.* **2016**, *2* (1), 1500292.

(12) Catalan, G.; Seidel, J.; Ramesh, R.; Scott, J. F. Domain Wall Nanoelectronics. *Rev. Mod. Phys.* **2012**, *84* (1), 119–156.

(13) Evans, D. M.; René S. D.; Holstad, T. S.; Erik V. P.; Mosberg, A. B.; Yan, Z.; Bourret, E.; Helvoort, A. T. J. V.; Selbach, S. M.; Meier, D. Observation of Electric-Field-Induced Structural Dislocations in a Ferroelectric Oxide. *Nano Lett.* **2021**. doi.org/10.1021/acs.nanolett.0c04816.

(14) Balke, N.; Winchester, B.; Ren, W.; Chu, Y. H.; Morozovska, A. N.; Eliseev, E. A.; Huijben, M.; Vasudevan, R. K.; Maksymovych, P.; Britson, J.; Jesse, S.; Kornev, I.; Ramesh, R.; Bellaiche, L.; Chen, L. Q.; Kalinin, S. V. Enhanced Electric Conductivity at Ferroelectric Vortex Cores in $BiFeO_3$. *Nat. Phys.* **2012**, *8* (1), 81–88.

(15) Wei, Y.; Matzen, S.; Maroutian, T.; Agnus, G.; Salverda, M.; Nukala, P.; Chen, Q.; Ye, J.; Lecoeur, P.; Noheda, B. Magnetic Tunnel Junctions Based on Ferroelectric $Hf_{0.5}Zr_{0.5}O_2$ Tunnel Barriers. *Phys. Rev. Appl.* **2019**, *12* (3), 031001.

(16) Aird, A.; Salje, E. K. H. Sheet Superconductivity in Twin Walls: Experimental Evidence of $WO_{3-x}$. *J. Phys. Condens. Matter* **1998**, *10* (22), L377.

(17) Rojac, T.; Ursic, H.; Bencan, A.; Malic, B.; Damjanovic, D. Mobile Domain Walls as a Bridge between Nanoscale Conductivity and Macroscopic Electromechanical Response. *Adv. Funct. Mater.* **2015**, *25* (14), 2099–2108.

(18) Agar, J. C.; Damodaran, A. R.; Okatan, M. B.; Kacher, J.; Gammer, C.; Vasudevan, R. K.; Pandya, S.; Dedon, L. R.; Mangalam, R. V. K.; Velarde, G. A.; Jesse, S.; Balke, N.; Minor, A. M.; Kalinin, S. V.; Martin, L. W. Highly Mobile Ferroelastic Domain Walls in Compositionally Graded Ferroelectric Thin Films. *Nat. Mater.* **2016**, *15* (5), 549–556.

(19) Sharma, P.; Zhang, Qi; Sando, Daniel; Lei, Chi Hou; Liu, Yunya; Li, Jiangyu; Nagarajan, Valanoor; Seidel, Jan. Nonvolatile Ferroelectric Domain Wall Memory. *Sci. Adv.* **2017**, *3* (6), 1–9.

(20) Farokhipoor, S.; Noheda, B.; Seidel, J.; Maksymovych, P.; Batra, Y.; Katan, A.; Yang, S. Y.; He, Q.; Baddorf, A. P.; Kalinin, S. V.; Yang, C. H.; Yang, J. C.; Chu, Y. H.; Salje, E. K. H.; Wormeester, H.; Salmeron, M.; Ramesh, R. Conduction through 71° domain Walls in $BiFeO_3$ Thin Films. *Phys. Rev. Lett.* **2011**, *107* (12), 3–6.

(21) Rojac, T.; Bencan, A.; Drazic, G.; Sakamoto, N.; Ursic, H.; Jancar, B.; Tavcar, G.; Makarovic, M.; Walker, J.; Malic, B.; Damjanovic, D. Domain-Wall Conduction in Ferroelectric $BiFeO_3$ Controlled by Accumulation of Charged Defects. *Nat. Mater.* **2017**, *16* (3), 322–327.

(22) Agarwal, R.; Sharma, Y.; Hong, S.; Katiyar, R. S. Modulation of Oxygen Vacancies Assisted Ferroelectric and Photovoltaic Properties of (Nd, V) Co-Doped $BiFeO_3$ Thin Films. *J. Phys. D: Appl. Phys.* **2018**, *51* (27), 275303.

(23) Liu, L.; Xu, K.; Li, Q.; Daniels, J.; Zhou, H.; Li, J.; Zhu, J.; Seidel, J.; Li, J.-F. Giant Domain Wall Conductivity in Self-Assembled $BiFeO_3$ Nanocrystals. *Adv. Funct. Mater.* **2021**, *31* (1), 2005876.

(24) Schaab, J; Skjærvø, S. H.; Krohns, S.; Dai, X.; Holtz, M. E.; Cano, A.; Lilienblum, M.; Yan, Z.; Bourret, E.; Muller, D. A.; Fiebig, M.; Selbach, S. M.; Meier, D. Electrical Half-Wave Rectification at Ferroelectric Domain Walls. *Nat. Nanotechnol.* **2018**, *13* (11), 1028–1034.

(25) Evans, Do. M.; Holstad, T. S.; Mosberg, A. B.; Småbråten, D. R.; Vullum, P. E.; Dadlani, A. L.; Shapovalov, K.; Yan, Z.; Bourret, E.; Gao, D.; Akola, J.; Torgersen, J.; Helvoort, A. T. J. V.; Selbach, S. M.; Meier, D. Conductivity Control via Minimally Invasive Anti-Frenkel Defects in a Functional Oxide. *Nat. Mater.* **2020**, *19* (11), 1195–1200.

(26) Werner, C. S.; Herr, S. J.; Buse, K.; Sturman, B.; Soergel, E.; Razzaghi, C.; Breunig, I. Large and Accessible Conductivity of Charged Domain Walls in Lithium Niobate. *Sci. Rep.* **2017**, *7* (1), 1–8.

(27) Schröder, M.; Haußmann, A.; Thiessen, A.; Soergel, E.; Woike, T.; Eng, L. M. Conducting Domain Walls in Lithium Niobate Single Crystals. *Adv. Funct. Mater.* **2012**, *22* (18), 3936–3944.

(28) Esin, A. A.; Akhmatkhanov, A. R.; Shur, V. Ya. Tilt Control of the Charged Domain Walls in Lithium Niobate. *Appl. Phys. Lett.* **2019**, *114* (9), 092901.

(29) Sluka, T.; Tagantsev, A. K.; Bednyakov, P.; Setter, N. Free-Electron Gas at Charged Domain Walls in Insulating $BaTiO_3$. *Nat. Commun.* **2013**, *4* (1808), 1–6.

(30) Guyonnet, J.; Gaponenko, I.; Gariglio, S.; Paruch, P. Conduction at Domain Walls in Insulating $Pb(Zr_{0.2}Ti_{0.8})O_3$ Thin Films. *Adv. Mater.* **2011**, *23* (45), 5377–5382.

(31) Tselev, A.; Yu, P.; Cao, Y.; Dedon, L. R.; Martin, L. W.; Kalinin, S. V.; Maksymovych, P. Microwave a.c. Conductivity of Domain Walls in Ferroelectric Thin Films. *Nat. Commun.* **2016**, *7* (1), 1–9.

(32) Xiao, H.; Wang, Y.; Jiao, N.; Guo, Y.; Dong, W.; Zhou, H.; Li, Q.; Sun, C. Understanding the Role of Oxygen Vacancy in Visible–Near-Infrared-Light-Absorbing Ferroelectric Perovskite Oxides Created by Off-Stoichiometry. *Adv. Electron. Mater.* **2019**, *5* (10), 1900407.

(33) Huang, J.; Chasteen, N. D.; Fitzgerald, J. J. X-Band EPR Studies of Ferroelectric Lead Titanate (PT), Piezoelectric Lead Magnesium Niobate (PMN), and PMN/PT Powders at 10 and 85 K. *Chem. Mater.* **1998**, *10* (12), 3848–3855.

(34) Maier, R. A.; Pomorski, T. A.; Lenahan, P. M.; Randall, C. A. Acceptor-Oxygen Vacancy Defect Dipoles and Fully Coordinated Defect Centers in a Ferroelectric Perovskite Lattice: Electron Paramagnetic Resonance Analysis of $Mn^{2+}$ in Single Crystal $BaTiO_3$. *J. Appl. Phys.* **2015**, *118* (16), 164102.

(35) Domingo, N.; Gaponenko, I.; Edwards, K. C.; Stucki, N.; Pérez,-D. V.; Escudero, C.; Pach, E.; Verdaguer, A.; Paruch, P. Surface Charged Species and Electrochemistry of Ferroelectric Thin Films. *Nanoscale* **2019**, *11* (38), 17920–17930.

(36) Domingo, N.; Pach, E.; Edwards, K. C.; Pérez,-D. V.; Escudero, Carlos; Verdaguer, Albert. Water Adsorption, Dissociation and Oxidation on $SrTiO_3$ and Ferroelectric Surfaces Revealed by Ambient Pressure X-Ray Photoelectron Spectroscopy. *Phys. Chem. Chem. Phys.* **2019**, *21* (9), 4920–4930.

(37) Park, C.; Chadi, D. Microscopic Study of Oxygen-Vacancy Defects in Ferroelectric Perovskites. *Phys. Rev. B* **1998**, *57* (22), R13961–R13964.

(38) Bednyakov, P. S.; Sturman, B. I.; Sluka, T. Tagantsev, A. K.; Yudin, P. V. Physics and Applications of Charged Domain Walls. *NPJ Comput. Mater.* **2018**, *4* (1), 1–11.

(39) Wu, W.; Horibe, Y.; Lee, N.; Cheong, S. W.; Guest, J. R. Conduction of Topologically Protected Charged Ferroelectric Domain Walls. *Phys. Rev. Lett.* **2012**, *108* (7), 077203.

(40) Gureev, M. Y.; Tagantsev, A. K.; Setter, N. Head-to-Head and Tail-to-Tail 180° Domain Walls in an Isolated Ferroelectric. *Phys. Rev. B* **2011**, *83* (18), 184104.

(41) Gong, J. J.; Li, C. F.; Zhang, Y.; Li, Y. Q.; Zheng, S. H.; Yang, K. L.; Huang, R. S.; Lin, L.; Yan, Z. B.; Liu, J. M. Interactions of Charged Domain Walls and Oxygen Vacancies in $BaTiO_3$: A First-Principles Study. *Mater. Today Phys.* **2018**, *6*, 9–21.

(42) Zhang, Y.; Lu, H.; Yan, X.; Cheng, X.; Xie, L.; Aoki, T.; Li, L.; Heikes, C.; Lau, S. P.; Schlom, D. G.; Chen, L.; Gruverman, A.; Pan, X. Intrinsic Conductance of Domain Walls in $BiFeO_3$. *Adv. Mater.* **2019**, *31* (36), 1902099.

(43) Seidel, J.; Martin, L. W.; He, Q.; Zhan, Q.; Chu, Y. H.; Rother, A.; Hawkridge, M. E.; Maksymovych, P.; Yu, P.; Gajek, M.; et al. Conduction at Domain Walls in Oxide Multiferroics. *Nat. Mater.* **2009**, *8* (3), 229–234.

(44) Campanini, M.; Gradauskaite, E.; Trassin, M.; Yi, D.; Yu, P.; Ramesh, R.; Erni, R.; Rossell, M. D. Imaging and Quantification of Charged Domain Walls in $BiFeO_3$. *Nanoscale* **2020**, *12* (16), 9186–9193.

(45) Geng, W. R.; Tian, X. H.; Jiang, Y. X.; Zhu, Y. L.; Tang, Y. L.; Wang, Y. J.; Zou, M. J.; Feng, Y. P.; Wu, B.; Hu, W. T.; Ma, X. L. Unveiling the Pinning Behavior of Charged Domain Walls in BiFeO3 Thin Films via Vacancy Defects. *Acta Mater.* **2020**, *186*, 68–76.

(46) Bencan, A.; Drazic, G.; Ursic, H.; Makarovic, M.; Komelj, M.; Rojac, T.. Domain-Wall Pinning and Defect Ordering in $BiFeO_3$ Probed on the Atomic and Nanoscale. *Nat. Commun.* **2020**, *11* (1), 1–9.

(47) Jang, J. H.; Kim, Y. M.; He, Q.; Mishra, R.; Qiao, L.; Biegalski, M. D.; Lupini, A. R.; Pantelides, S. T.; Pennycook, S. J.; Kalinin, S. V.; Borisevich, A. Y. In Situ Observation of Oxygen Vacancy Dynamics and Ordering in the Epitaxial $LaCoO_3$ System. *ACS Nano* **2017**, *11* (7), 6942–6949.

(48) Nukala, P.; Ahmadi, M.; Wei, Y.; Graaf, d. S.; Stylianidis, E.; Chakrabortty, T.; Matzen, S.; Zandbergen, H. W.; Björling, A.; Mannix, D.; Carbone, D.; Kooi, B.; Noheda, B. Reversible Oxygen Migration and Phase Transitions in Hafnia-Based Ferroelectric Devices. *Science.* **2021**, eabf3789.

(49) Paillard, C.; Geneste, G.; Bellaiche, L.; Dkhil, B. Vacancies and Holes in Bulk and at 180° Domain Walls in Lead Titanate. *J. Phys. Condens. Matter* **2017**, *29* (48), 485707.

(50) Barzilay, M.; Ivry, Y. Formation and Manipulation of Domain Walls with 2 nm Domain Periodicity in $BaTiO_3$ without Contact Electrodes. *Nanoscale* **2020**, *12* (20), 11136–11142.

(51) Elangovan, H.; Barzilay, M.; Seremi, S.; Cohen, N.; Jiang, Y.; Martin, L. W.; Ivry, Y. Giant Superelastic Piezoelectricity in Flexible Ferroelectric $BaTiO_3$ Membranes. *ACS Nano* **2020**, *14* (4), 5053–5060.

(52) Yücelen, E.; Lazić, I.; Bosch, E. G. T. Phase Contrast Scanning Transmission Electron Microscopy Imaging of Light and Heavy Atoms at the Limit of Contrast and Resolution. *Sci. Rep.* **2018**, *8* (1), 2676.

(53) Hawkes W. P., Eds. *Advances in Imaging and Electron Physics*; Elsevier, London, **2017**.

(54) Karim, H.; Delfin, D.; Chavez, L. A.; Delfin, L; Martinez, R.; Avila, J.; Rodriguez, C.; Rumpf, R. C.; Love, N.; Lin, Y. Metamaterial Based Passive Wireless Temperature Sensor. *Adv. Eng. Mater.* **2017**, *19* (5), 1600741.

(55) Barzilay, M.; Qiu, T.; Rappe, A. M.; Ivry, Y. Epitaxial $TiO_x$ Surface in Ferroelectric $BaTiO_3$: Native Structure and Dynamic Patterning at the Atomic Scale. *Adv. Funct. Mater.* **2020**, *30* (18), 1–9.

(56) Stemmer, S.; Höche, T.; Keding, R.; Rüssel, C.; Schneider, R.; Browning, N. D.; Streiffer, S. K.; Kleebe, H. J. Oxidation States of Titanium in Bulk Barium Titanates and in (100) Fiber-Textured $(Ba_xSr_{1-x})Ti_{1+y}O_{3+z}$ Thin Films. *Appl. Phys. Lett.* **2001**, *79* (19), 3149–3151.

(57) Torrisi, G.; Di Mauro, A.; Scuderi, M.; Nicotra, G.; Impellizzeri, G. Atomic Layer Deposition of $ZnO/TiO_2$ Multilayers: Towards the Understanding of Ti-Doping in ZnO Thin Films. *RSC Adv.* **2016**, *6* (91), 88886–88895.

(58) Potapov, P. L.; Jorissen, K.; Schryvers, D.; Lamoen, D. Effect of Charge Transfer on EELS Integrated Cross Sections in Mn and Ti Oxides. *Phys. Rev. B* **2004**, *70* (4), 045106.

(59) Li, C. -A.; Wu, S. -S. Topological States in Generalized Electric Quadrupole Insulators. *Phys. Rev. B* **2020**, *101* (19), 195309.

(60) Catalan, G.; Seidel, J.; Ramesh, R.; Scott, J. F. Domain Wall Nanoelectronics. *Rev. Mod. Phys.* **2012**, *84* (1), 119–156.

(61) Gao, W.; Addiego, C.; Wang, H.; Yan, X.; Hou, Y.; Ji, D.; Heikes, C.; Zhang, Y; Li, L.; Huyan, H.; Blum, T.; Aoki, T.; Nie, Y.; Schlom, D. G.; Wu, R.; Pan, X. Real-Space Charge-Density Imaging with Sub-Ångström Resolution by Four-Dimensional Electron Microscopy. *Nature* **2019**, *575* (7783), 480–484.

(62) Vanderlinde, J., Eds. *Classical Electromagnetic Theory*; Springer, Netherlands, **2005**.

(63) MacLaren, I.; Wang, L. Q.; McGrouther, D.; Craven, A. J.; McVitie, S.; Schierholz, R.; Kovács, A.; Barthel, J.; Dunin, -B. R. E. On the Origin of Differential Phase Contrast at a Locally Charged and Globally Charge-Compensated Domain Boundary in a Polar-Ordered Material. *Ultramicroscopy* **2015**, *154*, 57–63.

(64) Greenwood, N. N.; Earnshaw, A., Eds. *Chemistry of the Elements*; Elsevier, Oxford, **1997**.

(65) Qi, Y.; Liu, S.; Grinberg, I.; Rappe, A. M. Atomistic Description for Temperature-Driven Phase Transitions in $BaTiO_3$. *Phys. Rev. B* **2016**, *94* (13).

(66) Liu, S.; Grinberg, I.; Takenaka, H.; Rappe, A. M. Reinterpretation of the Bond-Valence Model with Bond-Order Formalism: An Improved Bond-Valence-Based Interatomic Potential for $PbTiO_3$. *Phys. Rev. B* **2013**, *88* (10), 104102.

(67) Liu, S.; Grinberg, I.; Rappe, A. M. Intrinsic Ferroelectric Switching from First Principles. *Nature* **2016**, *534* (7607), 360–363.

(68) Barzilay, M.; Elangovan, H.; Ivry, Y. Surface Nucleation of the Paraelectric Phase in Ferroelectric $BaTiO_3$: Atomic Scale Mapping. *ACS Appl. Electron. Mater.* **2019**, *1* (11),

2431–2436.

(69) Giannozzi, P.; Baroni, S.; Bonini, N.; Calandra, M.; Car, R.; Cavazzoni, C.; Ceresoli, D.; Chiarotti, G. L.; Cococcioni, M.; Dabo, I.; et al. QUANTUM ESPRESSO: A Modular and Open-Source Software Project for Quantum Simulations of Materials. *J. Phys. Condens. Matter* **2009**, *21* (39), 395502.

(70) Garrity, K. F.; Bennett, J. W.; Rabe, K. M.; Vanderbilt, D. Pseudopotentials for High-Throughput DFT Calculations. *Comput. Mater. Sci.* **2014**, *81*, 446–452.

# Supplementary Information for: *"Oxygen-vacancy-induced quasi-linear quadrupole at a locally charged ferroelectric domain wall"*

Hemaprabha Elangovan,[†,1,2] Maya Barzilay,[†,1,2] Jiawei Huang,[3,4,5] Shi Liu,[3,4,5] Shai Cohen[6] and Yachin Ivry[1,2,*]

[1] Department of Materials Science and Engineering, Technion – Israel Institute of Technology, Haifa 3200003, Israel.

[2] Solid State Institute, Technion – Israel Institute of Technology, Haifa 3200003, Israel.

[3] School of Science, Westlake University, Hangzhou, Zhejiang 310024, China.

[4] Institute of Natural Sciences, Westlake Institute for Advanced Study, Hangzhou, Zhejiang 310024, China.

[5] Key Laboratory for Quantum Materials of Zhejiang Province, Hangzhou Zhejiang 310024, China.

[6] Nuclear Research Centre-Negev, Beer-Sheva 84190, Israel.

[*]Correspondence to: <ivry@technion.ac.il>

[†] These authors contributed equally to the work.

**SI – Table of Contents**

**Large-scale micrographs**

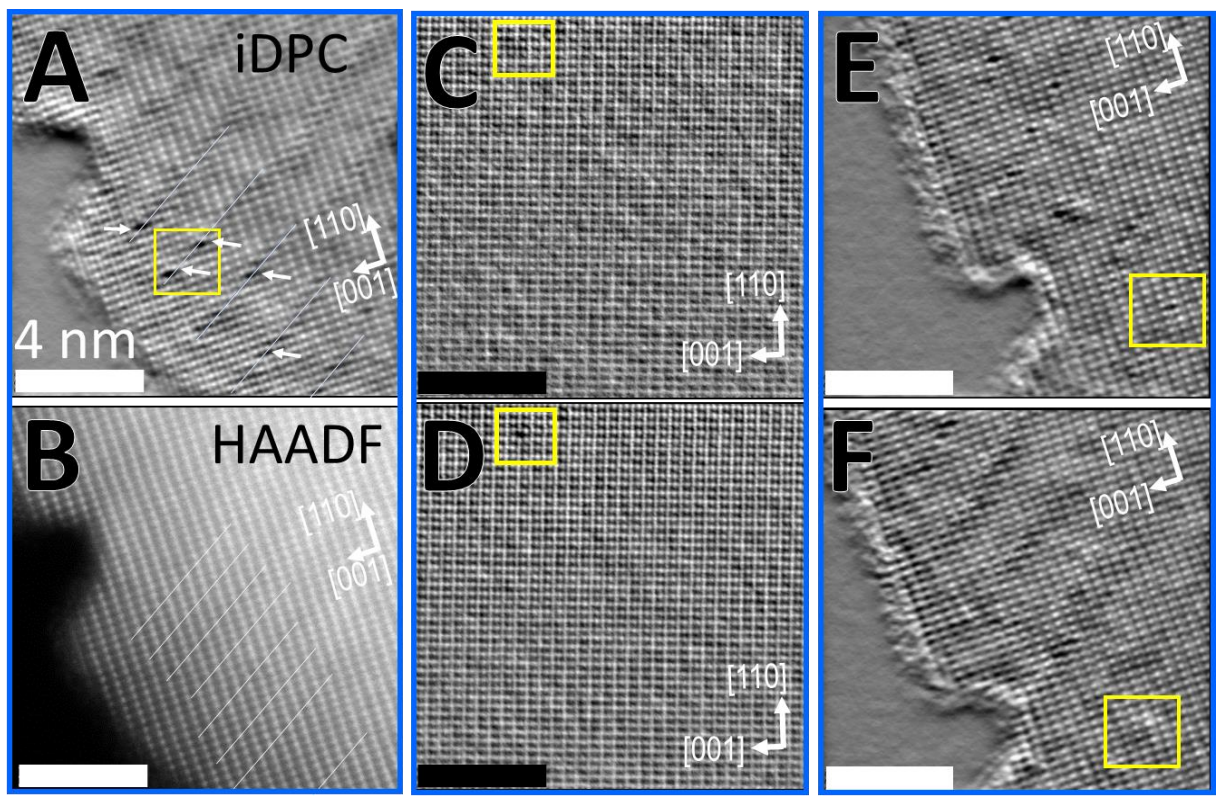

**Figure S1| Oxygen vacancy formation at 90° domain walls.** Simultaneously imaged (**A**) iDPC; (**B**) HAADF-STEM showing domain walls (highlighted) with oxygen vacancies of the area presented in Figure 1A-B. (**C**) iDPC image of the native structure of $BaTiO_3$, (**D**) iDPC micrograph of the same area as in (C) with an oxygen after an exposure to a dosage of 0.17 pA/pm$^2$, vacancy. (**E-F**) An iDPC micrograph at a different focal depth (2±1-nm difference) showing oxygen-vacancy localization. (C-F) are large-scale micrographs of the images presented in Figure 1C-F, respectively. All scale bars are 4 nm.

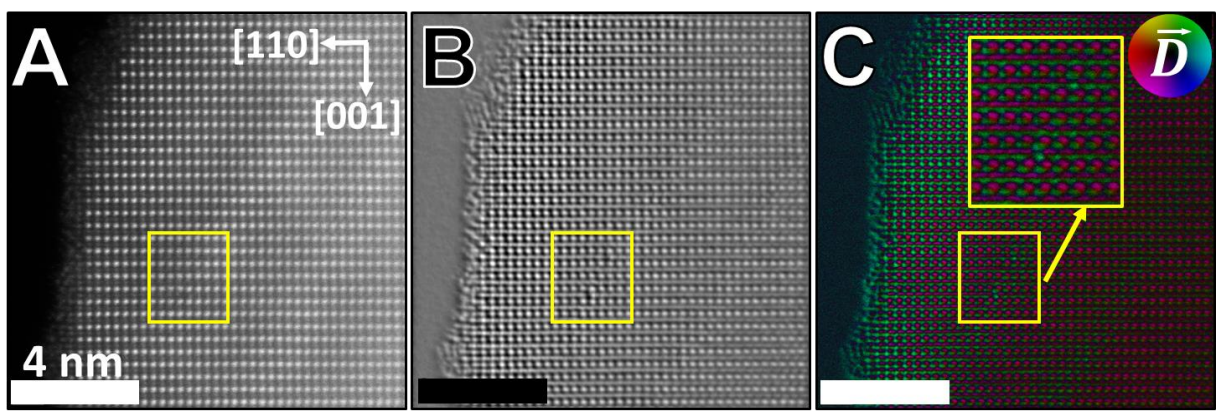

**Figure S2| Mechanical strain and dipole-moment distribution around an oxygen vacancy.** Simultaneously imaged (**A**) HAADF-STEM; (**B**) iDPC and (**C**) DPC micrographs, showing that two oxygen vacancies appear clearly at the iDPC and DPC images. The dipole distribution and the strain around these vacancies are presented in Figure 2A-B. The area shown in Figure 2A-B is highlighted in (B). Color wheel in (C) represents the orientation (color) and intensity (hue) of the electric-field displacement vector. All scale bars are 4 nm.

**Ba-Ba distance in a native BaTiO$_3$ structure**

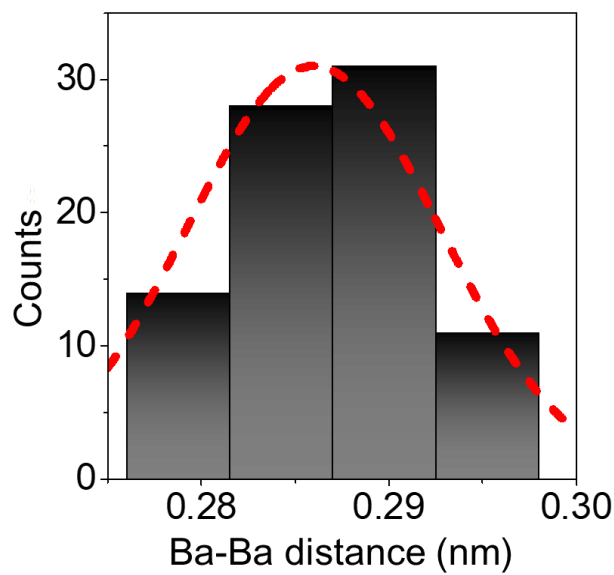

**Figure S3| Ba-Ba distance distribution within a pristine structure of BaTiO$_3$.** Ba-Ba distance along the [011] direction as was measured for >50 sites that do not contain nearby oxygen vacancies, showing 287±5 pm.

**EELS**

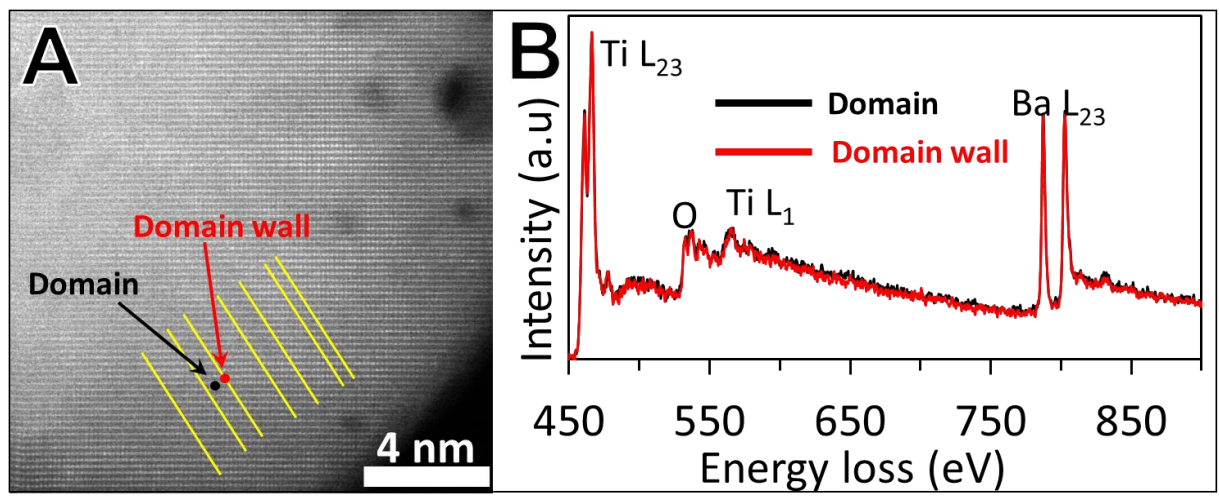

**Figure S4| EELS characterization across domain walls.** (**A**) An Annular Dark-Field (ADF) micrograph of an area within a BaTiO$_3$ crystallite with high-periodic domains.[50] (**B**) Large-are EELS spectra of data collected from within the domains (black) and the domain walls (red), indicating the energy position of Ba, Ti and O ions. The Ti L$_{23}$ is shown in Figure 2C. Each curve is an average of five different measurements along the line highlighted in (A).

**DPC-based charge-density calculation**

DPC is a STEM method, at which the electron-beam deflects due to local electromagnetic interactions with the sample, which exerts on it a Lorentz force. In the case of ferroelectrics, the deflection is vastly due to the contribution of bound charges to the displacement field $(\vec{D})$. Four sensitive detectors (A-D) that are organized together in the shape of a ring are awaiting to measure the electron beam after traveling through the sample. The deflection is correlated with the magnitude of $\vec{D}$ and maintains its orientation, allowing spatial mapping of the electromagnetic field within the material, following,

$$|\vec{D}| = \alpha \frac{I_{A-C}(x,y)\hat{i} + I_{B-D}(x,y)\hat{j}}{t\, I_{sum}(x,y)} \tag{S1}$$

where α is a calibration factor, $(x, y)$ is a coordinate at the sample plane, $I_{i-j}(x, y)$ is the intensity difference the signal collected at the i and j detector quadrants from the $(x, y)$ coordinate, $I_{sum}(x, y)$ is the total intensity at the four quadrants for the $(x, y)$ coordinate, *i.e.*, sum of the signals of all the four detectors and *t* is the sample thickness (~10 nm at the area of interest in this work).

The calibration factor corresponds primarily to the microscope and detector parameters:

$$\alpha = \left(\frac{R^2 - r^2}{R.C}\right)\left(\frac{mv}{e}\right) \tag{S2}$$

while *R* is the radius of the beam size at the detector plane (4 mm) and *r* represents the inner radius of the ring formed by the detectors (0.7 mm) at a camera length of *C* (295 mm). $e$, $m$, and $v$ are the electron charge and relativistic mass, and velocity, respectively for a given acceleration voltage (here, 200 kV). Micrographs of the individual quadrant detectors that were used to extract the displacement field and charge density distribution in Figure 3 are given in Figure S5.

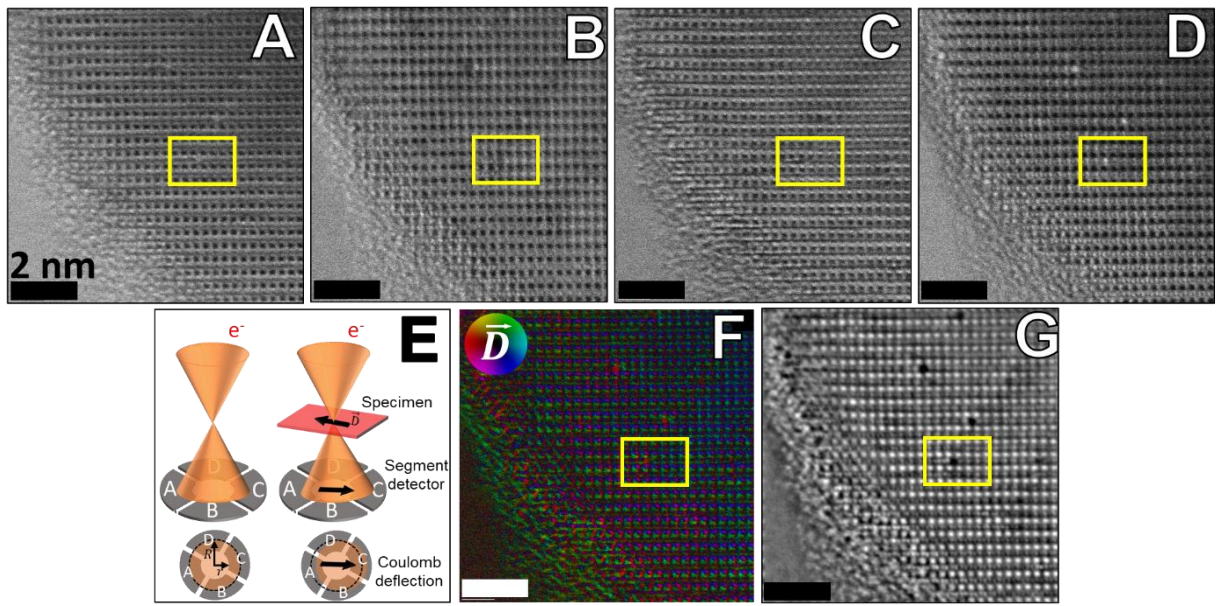

**Figure S5| Atomic-scale DPC segment-detector and iDPC mapping.** **(A-D)** Intensity signals collected at the respective A, B, C and D segments of the DPC detector. These signals were used to calculate the field and charge density in Figure 3, following Equations 1 and S2. (**E**) Simplified schematics of the electron beam deflection due to electric-field distribution in the specimen. (**F**) DPC and **(G)** iDPC images extracted from the segment-detectors signals (A-D, 0.5 s integration time) showing oxygen vacancies within domain walls. Yellow boxes denote the region of interest. Color wheel in (F) represents the orientation (color) and intensity (hue) of the electric-field displacement vector. All scale bars are 2 nm.

**Density of States (DOS) for PbTiO$_3$**

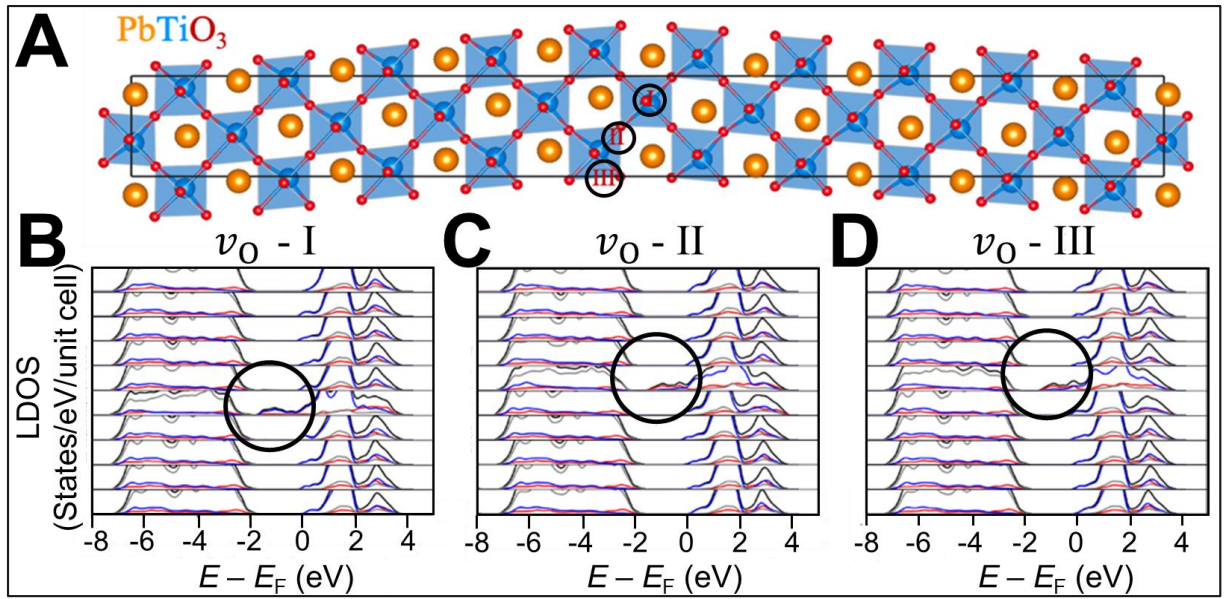

**Figure S6| DOS for PbTiO$_3$ with three differnet vacancies.** (**A**) The unit cell arrangement of PbTiO$_3$. (**B-D**) DOS for three different vacant sites, showing additional states in all sites, supporting the proposal of local conductivity. Vacant I and III are equatorial sites, while vacant II is the axial site in reference to the polarization direction.